\documentstyle[twocolumn,eqsecnum,aps]{revtex}
\begin{document}
\draft
\title{Finiteness Conditions for Light-Front Hamiltonians}
\author{M. Burkardt}
\address{Department of Physics\\
New Mexico State University\\
Las Cruces, NM 88003-0001\\U.S.A.}
\maketitle
\begin{abstract}
In the context of simple models, it is shown that demanding finiteness for 
physical masses with respect to a longitudinal cutoff, can be used to fix 
the ambiguity in the renormalization of fermions masses in the Hamiltonian 
light-front formulation. Difficulties that arise in applications of 
finiteness conditions to discrete light-cone quantization are discussed. 
\end{abstract}
\narrowtext
\section{Introduction}
Many advantages of the light-front (LF) formulation for bound state problems
arise from the manifest boost invariance in the longitudinal direction
\cite{all:lftd,world,brazil,dgr:elfe,mb:adv}.
The price for this advantage is that other symmetries, such as parity
or rotational invariance (for rotations around a transverse axis)
are no longer manifest \cite{mbrot,mb:parity}. 
From the technical point of view, the loss
of manifest parity and full rotational invariance implies that
LF Hamiltonians allow for a richer set of counter-terms in the 
renormalization procedure, i.e. in general LF Hamiltonians contain more 
parameters than the underlying Lagrangian.
 
Of course, even though parity and full rotational invariance are not
manifest symmetries in the LF formulation, a consistent calculation should
still give rise to physical observables which are consistent with these
symmetries. In Ref. \cite{mb:parity} this fact has been used to determine
one of these additional parameters by imposing parity covariance on
the vector form factor of mesons. While such a procedure is practical,
it is nevertheless desirable to have
alternative procedures available for determining these ``additional'' 
parameters in the Hamiltonian. 
In this paper, finiteness conditions are exploited to develop algorithms
for determining seemingly independent parameters in LF Hamiltonians.

As a specific example, let us consider a Yukawa model in 1+1 dimensions
\begin{equation}
{\cal L} = \bar{\psi}\left( i \not \;\!\!\!\partial 
-m-g\phi \right)\psi -\frac{1}{2}\phi\left( \Box +\lambda^2\right)\phi .
\end{equation}
In order to simplify the analysis further, we will in the following
consider the Yukawa model in a planar approximation (formally this can easily
be achieved by introducing ``color'' degrees of freedom and by assuming
an infinite number of ``colors''. However, while a planar approximation will
in the following always be implicitly used, explicit color degrees of freedom 
will not be shown in order to keep the notation simple.

The main difference between scalar and Dirac fields in the LF formulation is
that not all components of the Dirac field are dynamical: multiplying the
Dirac equation
\begin{equation}
\left( i\not \;\!\!\!\partial 
-m-g\phi \right)\psi =0
\end{equation}
by $\frac{1}{2}\gamma^+$ yields a constraint equation (i.e. an
``equation of motion'' without a time derivative)
\begin{equation}
i\partial_-\psi_{-}=\left(m+g\phi\right)\gamma^+\psi_{+}
,
\label{eq:constr}
\end{equation}
where
\begin{equation}
\psi_{\pm}\equiv \frac{1}{2}\gamma^\mp \gamma^\pm \psi .
\end{equation}
For the quantization procedure, it is convenient to eliminate
$\psi_{-}$ from the classical Lagrangian before imposing
quantization conditions, yielding
\begin{eqnarray}
{\cal L}&=&\sqrt{2}\psi_{+}^\dagger \partial_+ \psi_{+}
-\frac{1}{2}\phi\left( \Box +\lambda^2\right)\phi
-\psi^\dagger_{+}\frac{m^2}{\sqrt{2}i\partial_-}
\psi_{+}
\label{eq:lelim}
\\
&-&\psi^\dagger_{+}\left(
\phi
\frac{gm}{\sqrt{2}i\partial_-}
+\frac{gm}{\sqrt{2}i\partial_-}\phi\right)
\psi_{+}
-\psi^\dagger_{+}\phi\frac{g^2}{\sqrt{2}i\partial_-}
\phi\psi_{+} .
\nonumber
\end{eqnarray}
The rest of the quantization procedure very much resembles the procedure
for self-interacting scalar fields.
In particular, we must be careful about generalized tadpoles,
which might cause additional counter-terms in the LF Hamiltonian.
In the Yukawa model one usually (i.e. in a covariant formulation)
does not think about tadpoles. However, after eliminating $\psi_{(-)}$,
one is left with a four-point interaction in the Lagrangian, which does
give rise to time-ordered diagrams that resemble tadpole diagrams.
In fact, the four-point interaction gives rise to diagrams where
a fermion emits a boson, which may or may not self-interact, and then
re-absorb the boson at the same LF-time. \footnote{There are also tadpoles,
where the fermions get contracted. But those only give rise to an additional
boson mass counter-term, but not to the non-covariant fermion mass
counter-term that is investigated here.}
such interactions
cannot be generated by a LF Hamiltonian, i.e. the LF formalism
generally defines such tadpoles to be zero. An exception are the so-called
self-induced inertias, which arise from normal ordering the LF Hamiltonian.
These terms, which are ${\cal O}(g^2)$, are usually kept.

\section{Perturbative Counter-Term Analysis}
\begin{figure}
\unitlength1.cm
\begin{picture}(15,10.5)(0,1.5)
\includegraphics{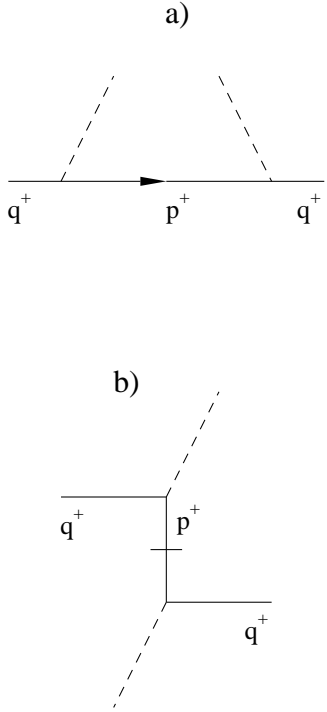}
\end{picture}
\caption{${\cal O}(g^2)$ contributions to the forward Compton amplitude.
(a) Intermediate fermion line on mass shell.
(b) Instantaneous fermion interaction contribution
(denoted by a slashed line).
}
\label{fig:compt2}
\end{figure}

At tree level, i.e. at order $g^0$, the kinetic mass and the vertex mass
have to be the same. In order to see this, let us consider the two
${\cal O}(g^2)$ Compton scattering diagrams in Fig. \ref{fig:compt2}. 
For simplicity we consider only forward scattering and we consider only 
diagrams which are singular.

The amplitude
with an on shell fermion intermediate state diverges as the $p^+$ momentum
of its intermediate fermion line goes to zero
\begin{equation}
T_o = \frac{g^2}{q^+-p^+}
\frac{\left(\frac{m}{q^+}+\frac{m}{p^+}\right)^2}
{q^- - \frac{m^2}{p^+} - \frac{\lambda^2}{q^+} }
\label{eq:to}
\end{equation}
(the subscript $o$ stands for on-shell).
This divergence is canceled exactly by the amplitude with an
instantaneous fermion line 
\begin{equation}
T_i = \frac{g^2}{q^+-p^+} \frac{1}{p^+} .
\end{equation}
The subscript $i$ stands for ``instantaneous''.
Note that this cancellation occurs if and only if the mass in the numerator
(the ``vertex mass'') and the mass in the denominator (the ``kinetic mass'')
are the same in Eq.~(\ref{eq:to}). This is also the only choice
of parameters that is consistent with parity invariance for Compton
scattering at ${\cal O}(g^2)$.

Choosing the vertex mass equal to the kinetic mass is also crucial for
a cancellation between the (momentum dependent!) self-induced inertia 
(kinetic mass) counter term \cite{pa:dlcq}
\begin{equation}
\Delta m^2 = \frac{g^2}{4\pi} \int_0^{p^+} \frac{dk^+}{k^+}
\label{eq:sii}
\end{equation} 
and the divergent piece of the ${\cal O}(g^2)$ self-energy
\begin{equation}
\Delta^{(2)} p^- = \frac{g^2}{4\pi} \int_0^{p^+} \frac{dk^+}{p^+-k^+}
\frac{ \left( \frac{m}{p^+}+\frac{m}{k^+} \right)^2}
{p^--\frac{m^2}{k^+} - \frac{\lambda^2}{p^+-k^+} }. 
\end{equation}
This well known result has recently also been obtained using so-called ladder
relations \cite{anton}, by investigating divergences in the
non-perturbative coupled Fock space equations for bound states.

While the self-induced inertia certainly cancels the divergent part of the
${\cal O}(g^2)$ self-energy, it has been questioned whether it also contains 
the correct finite part. In fact, in Ref. \cite{mb:parity}, parity invariance
for physical observables has been used to determine the finite piece of the
kinetic mass counter-term non-perturbatively.

However, the above analysis shows that the cancellation of divergences may
also be used to determine the finite piece: if the tree level cancellation
between instantaneous and on shell amplitudes is spoiled by a wrong choice
for the kinetic mass then higher order diagrams will contain a divergence
of integrals over longitudinal momenta as a result of the incomplete
cancellation. The question is --- and this will be subject of the rest of this
paper --- whether such ``finiteness conditions'' also arise at higher orders
in the coupling constants and whether they can be used to determine the finite
part of the kinetic mass counter-term.

\begin{figure}
\unitlength1.cm
\begin{picture}(15,9)(1.8,.5)
\includegraphics{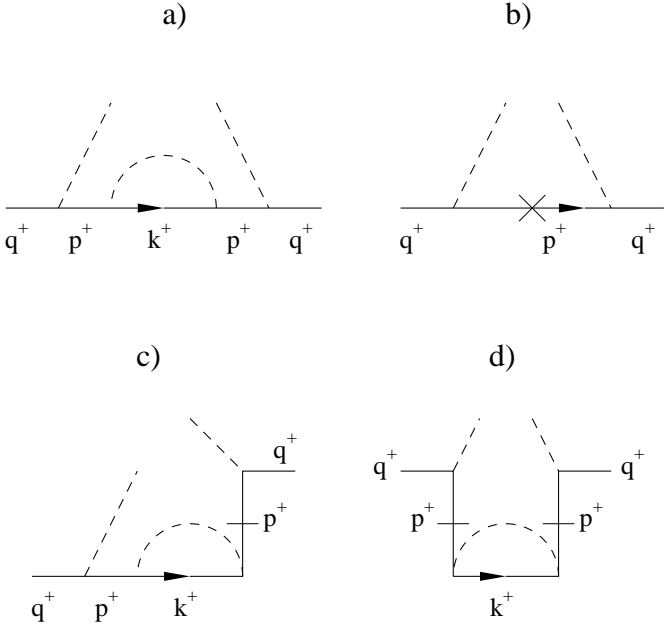}
\end{picture}
\caption{${\cal O}(g^4)$ contributions to the forward Compton amplitude.
(a) All fermion lines on mass shell.
(b) same as (a), but the loop replaced by the self-induced inertia.
(c) One of the two diagrams with an instantaneous fermion interaction 
(denoted by a slashed line) adjacent to the self-energy insertion.
(d) Both fermion propagators adjacent to the loop instantaneous.
}
\label{fig:compt4}
\end{figure}

For this purpose, let us consider the one-loop [${\cal O}(g^4)$] corrections
to the Compton amplitude. Again we restrict ourselves to planar diagrams.
Since we are interested only in corrections to the
$p^+\rightarrow 0$ singular contributions, it is also sufficient to consider
only loop corrections to the fermion line which propagates between the two
vertices. In LF-perturbation theory, we thus have to consider the four 
diagrams in Figure \ref{fig:compt4}.

\begin{figure}
\unitlength1.cm
\begin{picture}(15,4)(.75,-5.25)
\includegraphics{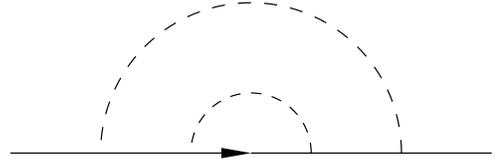}
\end{picture}
\caption{${\cal O}(g^4)$ contributions to the fermion 
self-energy, which is sensitive to the small $p^+$ behavior of the
${\cal O}(g^2)$ fermion self-energy.}
\label{fig:self4}
\end{figure}

Fig. \ref{fig:compt4} (a) and (b) together are finite (for finite $p^+$)
and contribute
\begin{equation}
T_{oo} = \frac{g^4\left(\frac{m}{q^+}+\frac{m}{p^+}\right)^2}
{4\pi\left(q^+-p^+\right)D_1^2}
\int_0^{p^+}\!\!\!\!\! dk^+ \left[
\frac{\left(\frac{m}{p^+}+\frac{m}{k^+}\right)^2}
{\left(p^+-k^+\right)D_2} + \frac{1}{k^+}\right],
\end{equation}
where
\begin{eqnarray}
D_1 &=& = q^--\frac{m^2}{p^+}-\frac{\lambda^2}{q^+-p^+}
\nonumber\\
D_2 &=& = p^--\frac{m^2}{k^+}-\frac{\lambda^2}{p^+-k^+}
\end{eqnarray}
(with $p^-\equiv q^--\frac{\lambda^2}{q^+-p^+}$)
are the energy denominators for the intermediate states.
The diagrams with one or two instantaneous lines are finite without
counter-terms (for finite $p^+$) and yield, respectively 
\begin{eqnarray}
T_{oi} &=& \frac{2g^4\left(\frac{m}{q^+}+\frac{m}{p^+}\right)}
{4\pi\left(q^+-p^+\right)p^+D_1}
\int_0^{p^+}\!\!\!\!\! dk^+ \frac{\left(\frac{m}{p^+}+\frac{m}{k^+}\right)}
{\left(p^+-k^+\right)D_2}
\nonumber\\
T_{ii} &=& \frac{g^4}{\left(q^+-p^+\right){p^+}^2}
\int_0^{p^+}\!\!\!\!\! dk^+ \frac{1}
{4\pi\left(p^+-k^+\right)D_2} .
\end{eqnarray}
All three amplitudes diverge like $1 /p^+$ as $p^+ \rightarrow 0$ !
One finds
\begin{eqnarray}
\lim_{p^+\rightarrow 0} p^+T_{oo}&=& \frac{g^4}{4\pi q^+}\left[\frac{1}{m^2}\ln 
\frac{\lambda^2}{m^2}
-\int_0^1 dx \frac{2+x}{m^2\left(1-x\right)+\lambda^2x}\right]
\nonumber\\
\lim_{p^+\rightarrow 0} p^+T_{oi}&=&\frac{g^4}{4\pi q^+}
\int_0^1 dx \frac{2+2x}{m^2\left(1-x\right)+\lambda^2x}
\nonumber\\
\lim_{p^+\rightarrow 0} p^+T_{ii}&=& -\frac{g^4}{4\pi q^+}
\int_0^1 dx \frac{x}{m^2\left(1-x\right)+\lambda^2x}
.
\label{eq:tlimit}
\end{eqnarray}
The divergence at small
$p^+$ does {\it not} cancel when one sums up the three terms. \footnote{An
exception is the ``supersymmetric'' case $m^2=\lambda^2$.}
In fact, what one finds is
\begin{equation}
\lim_{p^+\rightarrow 0} p^+\left(T_{oo}+T_{oi}+T_{ii}\right)
= \frac{g^4}{4\pi m^2q^+} \ln \frac{\lambda^2}{m^2}.
\end{equation}
Since there are no diagrams other than the ones listed in Fig.~\ref{fig:compt4}
which are singular at ${\cal O}(g^4)$, this implies that there is a problem:
The ${\cal O}(g^4)$ self-energy of a fermion (Fig.~\ref{fig:self4})
is obtained by integrating the ${\cal O}(g^4)$ forward Compton amplitude
over $p^+$ and one obtains a logarithmic divergence!
This divergence should not be there since Yukawa$_{1+1}$ is 
super-renormalizable. Already in perturbation theory,
the Yukawa model on the LF with only the self-induced
inertias added as counter-terms does not lead to finite answers.
Surprisingly, the resolution to this problem does {\it not} require to add
another infinite counter-term. In Ref. \cite{mb:parity} a finite kinetic
mass counter-term (in addition to the infinite self-induced inertias) was
introduced and it was found to be necessary in order to obtain
parity invariant form-factors. The effect of a ${\cal O}(g^2)$ kinetic mass 
counter-term is an additional ${\cal O}(g^4)$ term
in the forward Compton amplitude
\begin{equation}
T_{\Delta m^2} = \frac{g^2 \left(\frac{m}{q^+}+\frac{m}{p^+}\right)^2}
{\left(q^+-p^+\right)D_1^2} \frac{\Delta m^2_{kin}}{p^+} .
\end{equation}
It can easily be verified that the choice
\begin{equation}
\Delta m^2_{kin} = \frac{g^2}{4\pi} \ln \frac{m^2}{\lambda^2}
\label{eq:deltam}
\end{equation}
leads to
\begin{equation}
\lim_{p^+\rightarrow 0} p^+\left(T_{oo}+T_{oi}+T_{ii}+T_{\Delta m^2}\right)
=0
\end{equation}
and hence the ${\cal O}(g^4)$ self-energy of a fermion is finite with this
(and only this) particular choice for the kinetic mass counter-term.
Note that exactly the same values for the ${\cal O}(g^2)$ kinetic mass 
counter-term also lead to parity invariant scattering amplitudes.

Note that while the calculations presented above had been done for a
scalar Yukawa theory, very similar results hold for models with similar
interactions, such as pseudoscalar Yukawa of fermions coupled to the $\perp$
component of a vector field.

\section{A Non-Perturbative Example}

For a non-perturbative example, let us consider the model introduced
in Ref. \cite{mb:hala}: \footnote{For details and definitions the reader is
referred to this paper.} fermions in 3+1 dimensions coupled to the
$\perp$ components of a massive vector field in planar approximation.

The non-perturbative Green's function for a
fermion in this model can be written in the form
\begin{eqnarray}
G(p^\mu) &=& \gamma^+ p^-G_+(2p^+p^-,{\vec p}_\perp^2)
+\gamma^- p^+G_-(2p^+p^-,{\vec p}_\perp^2)
\nonumber\\
& &+ \not \;\!\!\!k_\perp G_\perp(2p^+p^-,{\vec p}_\perp^2)+
G_0(2p^+p^-,{\vec p}_\perp^2),
\label{eq:lf1}
\end{eqnarray}
where each of the $G_i$ has a spectral representation
\begin{eqnarray}
G_i(2p^+p^-,{\vec p}_\perp^2) = \int_0^\infty dM^2
\frac{\rho_i^{LF}(M^2,{\vec p}_\perp^2)}
{2p^+p^--M^2+i\varepsilon}.
\label{eq:speclf}
\end{eqnarray}
Note that $\mbox{tr}\left(\gamma^- G\right)$ cannot contain a term
proportional to
$\frac{1}{p^+} \int dM^2 \rho (M^2,p_\perp^2 )/
(2p^+p^- -M^2+i\varepsilon)$ because this would lead to severe small $p^+$
divergences, which are not canceled by the self-induced 
inertias. \footnote{Below this assumption will be shown to be self-consistent.}

From the fermion Green's function, one computes the self-energy
self-consistently via
\begin{equation}
G^{-1}=\not \;\!\!\! p -m-\Sigma,
\end{equation}
where
\begin{equation}
\Sigma=\gamma^+\Sigma_++\gamma^-\Sigma_-+\Sigma_0.
\end{equation}
For the LF components of the self-energy one finds
\begin{equation}
\Sigma_i=g^2 \int_0^\infty dM^2 \int_0^{p^+} \frac{dk^+}{k^+}
\int \frac{d^2k_\perp}{16\pi^3} f_i,
\end{equation}
where
\begin{eqnarray}
f_+&=& \left[\frac{\tilde{k^-}}
{D \left(p^+-k^+\right)} -\frac{1}{p^+} \right]
\rho_+\left(M^2,{\vec k}_\perp^2\right)\nonumber\\
f_-&=& \frac{k^+\rho_-\left(M^2,{\vec k}_\perp^2\right)}
{D \left(p^+-k^+\right)} \nonumber\\
f_+&=& -\frac{\rho_0\left(M^2,{\vec k}_\perp^2\right)}
{D \left(p^+-k^+\right)} 
\end{eqnarray}
and
\begin{eqnarray}
D&=&p^--\frac{M^2}{2k^+} - \frac{\lambda^2+\left({\vec p}_\perp
-{\vec k}_\perp \right)^2}{2\left(p^+-k^+\right)} \nonumber\\
\tilde{k}^-&=&p^-- \frac{\lambda^2+\left({\vec p}_\perp
-{\vec k}_\perp \right)^2}{2\left(p^+-k^+\right)} .
\end{eqnarray}

Note that in this toy model from Ref. \cite{mb:hala}, such a 
truncation of the Schwinger-Dyson equations is exact.

In order to be able to investigate whether the self-induced inertias
cancel the infinite part of the self-energy one needs to know the small $p^+$
behavior of $G$ and thus the small $p^+$ behavior of $\Sigma$.

As an example, let us suppose that $\Sigma = c\gamma^+/2p^+$, in
which case one finds for $p^+\rightarrow 0$
\begin{equation}
G = \frac{1}{\not \;\!\!\! p-m-\Sigma}
\longrightarrow -\frac{c}{c+m^2+{\vec p}_\perp^2}
\frac{\gamma^+}{p^+},
\end{equation}
which diverges, while the propagator for $c=0$
remains finite in this limit. The self induced inertias cancel the 
infinite part of the self-energy in the case where the fermion propagator 
inside the loop is a free propagator. If one wants that the same cancellation
occurs with the full propagator, it is thus necessary that the self-energy
which  modifies the propagator remains finite as $p^+\rightarrow 0$.

We will now investigate the consequences of this fact for the model
studied in Ref. \cite{mb:hala}. In particular, we will focus on the
$\gamma^+$ component of $\Sigma$, which is the most singular term as 
$p^+\rightarrow 0$. Including only the self-induced inertia counter-term,
one finds \cite{mb:hala}
\begin{eqnarray}
\Sigma^+ &=& g^2 p^-\int_0^{p^+} \frac{dk^+}{p^+} 
\int \frac{d^2k_\perp}{8\pi^3}
G_+(2k^+\tilde{k}^-,{\vec k}_\perp^2) \nonumber\\
& &-\frac{g^2}{2p^+} \int_0^\infty \!\!dM^2\!
\int \frac{d^2k_\perp}{8\pi^3}
\rho_+(M^2,{\vec k}_\perp^2) \ln \frac{M^2}{\lambda^2+{\vec k}_\perp^2}.
\label{eq:s+}
\end{eqnarray}
The first term on the r.h.s. of Eq. (\ref{eq:s+}) is finite as 
$p^+\rightarrow 0$, but the second term diverges in this limit. 
This second term is the non-perturbative
analog of the $\ln \lambda^2/m^2$ term in the one-loop self-energy, which 
we had to cancel by introducing a finite kinetic mass counter-term
in order to avoid divergences in the two-loop self-energy.
Here we also need to cancel the second term 
by means of a kinetic mass counter-term in order to obtain finite
solutions to the self-consistent LF version of the Schwinger-Dyson equations.

In summary, what we have found is that the two-loop result generalizes
directly to all orders in this non-perturbative example. In fact, one can
show that the result generalizes to an entire class of models with
Yukawa (scalar and pseudo-scalar) interactions as well as models with
couplings to transverse components of vector fields.
However, I was not able to show that the result generalizes to all orders
to models with couplings to longitudinal components of a vector field
(i.e. gauge theories). Semi-perturbative considerations suggest that
the results also apply to dimensionally reduced models for QCD \cite{anton}
as well as $\perp$ lattice QCD, but I could not find a general proof
(beyond perturbative calculations). Nevertheless, let us in the following 
conjecture that the result generalizes at least to models for QCD which have 
only a two-dimensional continuum (such as dimensionally reduced models and 
the $\perp$ lattice) and let us discuss the consequences.

First of all, at least in principle, this means that one can use the
dependence of physical masses on the longitudinal cutoff to fine-tune 
the finite part of the kinetic mass counter term. However, finite 
physical masses do not necessarily imply that one has the correct
kinetic mass. To illustrate this point, consider simple quantum mechanical
scattering in two or more space dimensions from a $\delta$-function potential.
Regardless of the sign of the potential, higher order Born terms in the
scattering amplitude all diverge. Nevertheless, non-perturbative
energies for physical states
only diverge in the attractive case, but not in the repulsive case.
What this implies for the fine-tuning procedure of the kinetic mass 
in the LF-Yukawa model is the following: 
If one wants to use finiteness conditions to determine the
correct value of the kinetic mass term, one needs to vary the
physical mass and study the cutoff dependence of physical masses
for each kinetic mass. If the kinetic mass is smaller than the correct value,
physical masses will become tachyonic as the cutoff is removed.
On the other hand, for any value of the kinetic mass which is larger than 
the correct value the spectrum will {\it not} necessarily diverge.
\footnote{One can construct examples where the whole spectrum diverges
as a divergent positive term is added to the Hamiltonian, but also
examples where part of the spectrum remains finite.}
The correct kinetic mass is thus obtained by working right at 
(i.e. infinitesimally above) the critical point where the spectrum becomes 
tachyonic!
\footnote{Working near a critical point is a frightening prospect for
practitioners, but the critical point is only of first order.}

While this algorithm seems to be quite easy to use, there are several
reasons why one should be very careful in its application to practical
problems.

First of all, in many LF calculations higher Fock components typically
contribute only small corrections to physical masses at a given cutoff.
What this means for the practical applicability of finiteness criteria
is that any $\log$-dependencies on a cutoff (which one needs to identify
in order to apply finiteness conditions)  may enter with a very
small coefficient so that they might be practically invisible. 

Secondly, it is very important to
discuss the cutoff scheme dependence! So far, we have on purpose
avoided to specify a cutoff procedure --- which one always has to
do when dealing with divergent (or potentially divergent) quantities.
The reason we did not have to specify the cutoff procedure is that
the one-loop divergence is canceled locally (the singularities of
the integrand cancel) by the self-induced inertia and higher order
divergences are also canceled locally by the finite kinetic mass
counter terms. However, we still assumed implicitly that
the result for the inner loop was (apart from trivial kinematical
factors) momentum independent --- otherwise it would not have been
sufficient to add merely a number (not a function) as a counter-term. 

It is easily possible to introduce cutoffs which have this property,
for example an invariant mass difference cutoff at each 3-point
vertex and a cutoff for the instantaneous fermion exchange diagrams which
is consistent with cutoffs on iterated 3-point vertices.
However, one of the most popular cutoffs used in non-perturbative 
LF-calculations
is DLCQ, where all momenta are discretized and thus a cutoff on the
longitudinal momenta is provided
by the spacing of the grid in momentum space. With such a cutoff procedure 
the self energy of a fermion does depend on its momentum (beyond the trivial
$1/p^+$ dependence). This point will be elaborated in Section \ref{sec:dlcq}.
However, before we discuss numerical implications in DLCQ, let us first 
consider finiteness relations derived by using perturbative relations between
Fock space components in non-perturbative bound state problems.

\section{Finiteness conditions and ladder relations}
\label{sec:ladder}
In bound state problems it is often possible to relate Fock space
components which are highly off energy shell to lower Fock components
using perturbation theory. This fact has been used within a dimensionally
reduced model for QCD in Ref. \cite{anton}
to relate the end-point behavior of Fock space amplitudes with n+1
quanta to Fock space amplitudes with n quanta, via \footnote{No distinction
between vertex and kinetic masses has been made in Ref. \cite{anton}.}
\begin{equation}
\psi_{n+1}(x_1,x_2,...,x_{n-1},0) \propto \frac{1}{m\sqrt{x_{n-1}}}
\psi_{n}(x_1,x_2,...,x_{n-1}).
\label{eq:ladder}
\end{equation} 
Eq. (\ref{eq:ladder}) shows that wave functions in higher Fock
components do not
vanish near the end-point (i.e. for vanishing fermion momenta), which
leads to divergent matrix elements  of the kinetic energy as well
as the interaction. The divergence that arises when only the
fermion momentum goes to zero is canceled exactly by the self-induced
inertias [Eq. (\ref{eq:sii})] if and only if the vertex mass $m_V$
and the kinetic mass $m_{kin}$
are the same. 

In Ref. \cite{anton} it is thus claimed that the bound state equation 
(with $m_V=m_{kin}$) is finite.
This claim is false: the Hamiltonian studied in Ref. \cite{anton}
is in general not finite! The point is that both Eq. (\ref{eq:ladder})
as well as the cancellation conditions require to be modified when
two momenta go to zero simultaneously. The best way to see that
without going into too much detail is to consider the matrix element
which connects states which differ by one boson. Such a matrix
element involves the inverse of the momenta of both the incoming and
outgoing fermion. If only the outgoing momentum goes to zero, then
the term with the inverse of the momentum of the incoming fermion
can obviously be neglected. However, this is not the case if both incoming
and outgoing momentum go to zero simultaneously. Since the vanishing of both
incoming and outgoing fermion momenta also implies that
the momentum of the
emitted boson also vanishes, one can therefore conclude that the end-point
behavior gets modified if the momenta of both the fermion and a
boson vanish simultaneously.

Furthermore, the cancellation conditions also get modified when
proper care is taken for the case where several momenta vanish
simultaneously. In particular, in order for the Hamiltonian to
give finite results one does in general need to keep $m_V \neq m_{kin}$.

The two-loop example considered above can be considered a formal
proof (by counter-example!) for these intuitively obvious facts.

In Ref. \cite{anton}, numerical evidence is offered for the
finiteness claim made in the same paper. Below, in Section
\ref{sec:dlcq}, it will be demonstrated that the (logarithmic)
divergence arising from the two-loop diagram shows up only
for very large values of the DLCQ parameter $K$. This is probably
the main reason why the divergence did not show up in the
numerical results presented in Ref. \cite{anton}.

\section{Finiteness conditions in DLCQ} 
\label{sec:dlcq}

It is very easy to see that discretization in momentum space leads
to a momentum dependent self-mass. Compared to a continuum calculation,
integrals are approximated by sums and the number of points over which
the summation is performed is determined by the total momentum.
In this section, we will investigate the implications of this obvious fact
for finiteness conditions.

In order to simplify the discussion, let us consider a cutoff which is
very similar to the DLCQ cutoff, namely a sharp momentum cutoff (in the
continuum) on all momenta that are smaller than an arbitrary
constant $\varepsilon$.

The point is that since the cutoff acts both on the boson and on the
fermion line, self-energy corrections to the ${\cal O}(g^4)$
Compton amplitude are absent for $p^+<2\varepsilon$ and they are suppressed
for $p^+$ near that value. On the other hand, a 
(momentum independent!) kinetic mass counter-term would contribute all
the way down to the cutoff, namely $p^+=\varepsilon$.
For the self-energy this implies that there is an incomplete
cancellation between terms that would cancel if the cutoff on the
inner loop would be sent to zero {\it before} the outer loop integration is
performed.

In order to illustrate what consequences this might have, let us consider a 
simple mathematical model which has the right qualitative features:
let us assume that the sum of amplitudes in Fig. \ref{fig:compt4}
in the presence of a cutoff is given by
\begin{equation}
p^+T=\frac{c}{q^+}\Theta (p^+-2\varepsilon).
\end{equation}
Including a kinetic mass counter-term $\Delta m^2_{kin}$, the two loop
self-energy is then given by
\begin{equation}
\Delta^{(4)} q^- \propto \int_\varepsilon^{q^+} \frac{dp^+}{p^+}
\left[c\Theta (p^+-2\varepsilon)-\Delta m^2_{kin}\right].
\end{equation}
Despite the fact that the integral over the self-energy piece starts at
$p^+=2\varepsilon$, while the integral over the mass counter-term contribution
starts at $p^+=\varepsilon$, the unique choice for $\Delta m^2_{kin}$ which
yields a finite two loop self-energy as $\varepsilon \rightarrow 0$ is
$\Delta m^2_{kin} =c$. And the result of the integral in this case
is $-c\ln 2$ (independent of $\varepsilon$). 
Had we taken the limit $\varepsilon \rightarrow 0$ in the
integrand, then the integrand would identically vanish and the integral would
be zero. In other words, the finiteness condition would have given us the
correct value for the kinetic mass counter-term at ${\cal O}(g^2)$,
but the wrong result for the physical mass at  ${\cal O}(g^4)$.

\begin{figure}
\unitlength1.cm
\begin{picture}(15,10)(1.25,1)
\includegraphics{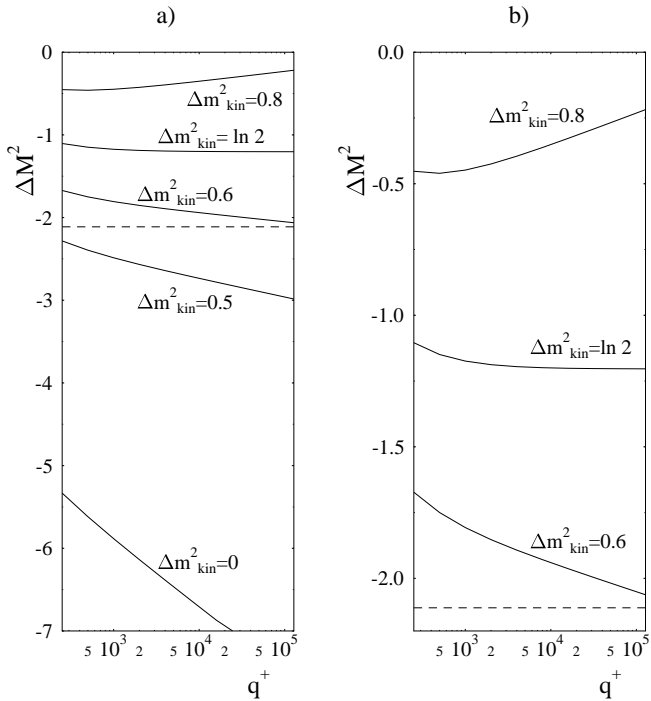}
\end{picture}
\caption{(a) Two loop self energy of a fermion calculated using DLCQ (with
anti-periodic boundary conditions for the fermion) as a function of
the momentum of the fermion. The different curves represent different
kinetic mass counter-terms $\Delta m^2_{kin}$. (b) Same as (a) but a smaller 
interval of the y-axis is shown to simplify determination of the kinetic mass
counter-term for which the result is stable for large momenta.
The covariant result is indicated as a dashed line.
Note that, while $\Delta m^2_{kin}=\ln 2$ leads to a convergent result,
it does not converge towards the covariant answer.
}
\label{fig:2_1}
\end{figure}

In order to demonstrate that this problem does indeed occur in DLCQ,
let us consider a concrete problem, namely the  ${\cal O}(g^4)$
self-mass $\Delta M^2 \equiv q^+ \delta^{(4)} q^-$ 
resulting from the rainbow diagram (Fig. \ref{fig:self4}).
Even though we know the correct kinetic mass counter-term for this case
from Eq. (\ref{eq:deltam}), let us pretend here that we do not know it
and let us consider the two loop self-energy both as a function of the
momentum $q^+$ (in discrete units) and the kinetic mass counter-term.
The coupling constant is set to $g=\sqrt{4\pi}$, and for the masses
we choose $\lambda^2=1$ and $m^2=2$. Figure \ref{fig:2_1} shows
$4 \pi q^+$ times the self-energy (including the kinetic mass counter-term)
of the fermion as a function of
$q^+$ for different values of the parameter $\Delta m^2_{kin}$.
There are several things one can learn from this calculation.
First of all, Fig. \ref{fig:2_1} clearly shows that a kinetic mass 
counter-term $\Delta m^2_{kin}$
(in addition to the self-induced inertia) is necessary
in order to obtain finite results: the two-loop result for $\Delta M^2$  
obviously diverges when one sets $\Delta m^2_{kin}=0$. 

Secondly, the procedure is not very sensitive since the divergence
is only logarithmic and the coefficient of the divergent piece is not
very large. In order to obtain a precise picture about which 
value for the kinetic mass parameter 
leads to a convergent one has to go to values of $q^+>1000$,
which is forbiddingly large for a non-perturbative calculation, but
a reasonable estimate can already be obtained at lower values.

Thirdly, the finiteness condition does give the correct value
for the kinetic mass counter-term. Only for $\delta m^2_{kin}\approx \ln 2$
(for $m^2=2$ and $\lambda^2=1$ one finds no noticeable $q^+$ dependence
of the self-energy for large $q^+$. Even small deviations lead to
a $\log q^+$ divergence proportional to that deviation.

Finally, and this is very important, despite the fact that the finiteness
condition yields the correct value for $\delta m^2_{kin}$, the final
result of the ${\cal O}(g^4)$ differs from the covariant result:
For  $\Delta m^2_{kin}\approx \ln 2$ one finds
$\lim_{q^+\rightarrow \infty} \Delta M^2 \approx -1.204$, while the correct
(covariant) result for the two loop diagram (Fig. \ref{fig:self4})
is given by $\Delta M^2 \approx -2.112$ for the same masses and couplings.
As we discussed above, this is because in DLCQ the momentum
of a line that enters a sub-loop is not necessarily high above the cutoff
inside that sub-loop. Therefore, the sensitivity to the cutoff never 
goes away --- not even when the overall momentum is sent to infinity.
Another way to look at this result is to conclude that in DLCQ one
cannot introduce just one kinetic mass counter-term, but instead one
needs to introduce a kinetic mass which depends on on the momentum.
Formally, this should not come as a surprise, since the boost invariance
(which is normally manifest in LF quantization) is broken by the DLCQ
regulator \cite{all:lftd}. However, in a number of examples, such as
1+1 dimensional QED/QCD and theories with only self-interacting scalar fields, 
momentum dependent counter terms are not
necessary and DLCQ workers have become accustomed to assume momentum
independence of all counter-terms as a starting point.
Unfortunately, the Yukawa model that we have considered here is a clear 
counter-example to this simplified picture.

Of course, for a perturbative diagram one can always calculate the proper 
momentum dependence, but this seems impossible to do analytically in a
non-perturbative context. An alternative procedure is the one employed
in Refs. \cite{mb:parity,mb:hala}, where a momentum dependent kinetic
mass is introduced such that the physical mass of the lightest states
is independent of the momentum. The physical mass then replaces the bare
kinetic mass as a renormalization parameter. 
In Refs. \cite{mb:parity,mb:hala} the new parameters were determined by 
imposing parity invariance on physical amplitudes or by comparison with
a covariant calculation. However, it is not obvious how to translate the
finiteness condition for kinetic masses into a condition for
the physical masses.

The fact that a simple (i.e. momentum independent) kinetic mass counter-term
yields incorrect results also means that the ansatz for the LF Hamiltonian
in theories with fermions and Yukawa type interactions 
(this includes QED/QCD!)
used by DLCQ workers (see for example Refs. \cite{pa:dlcq,anton}) is
insufficient.

There are several obvious patches that one can apply to the DLCQ calculations,
but they all seem to have one feature in common: one needs to introduce
another cutoff --- beyond DLCQ --- which has the feature that it gives
momentum independent results. Typical examples are a Pauli-Villars regulator
\cite{mbrot,john} or a cutoff on the invariant energy transfer.
Of course, even with a cutoff that gives momentum independent results, 
one still
needs to keep the kinetic mass as an ``independent parameter'', 
\footnote{An exception is Pauli-Villars regularization with sufficiently
many regulator particles \cite{mbrot,john}.} which
then needs to be determined using for example parity or finiteness conditions,
but at least one does not have to introduce a kinetic mass which is a function
of the momentum.
It is not clear whether adding an ${\cal O}(g^4)$ kinetic mass counterterm 
to correct for the artefacts introduced by the DLCQ cutoff leads to a
consistent procedure at  ${\cal O}(g^6)$ or higher.

\section{Summary}
We have investigated the conditions under which light-front Hamiltonians with
fermions interacting via Yukawa type interactions 
(including interactions to the transverse component of a vector field)
lead to convergent loop integrals
at small values of the LF momentum $p^+\equiv p^0+p^3$. In the continuum,
it was found that it is both necessary and sufficient to add a kinetic 
mass counter-term (in addition
to the self-induced inertias) to the Hamiltonian in order to obtain
finite results w.r.t. the small $p^+$ cutoff for higher order diagrams. 
That additional parameter is
determined by demanding finiteness for the $p^+$ integrals.
Imposing such a finiteness condition makes sense, since the small 
$p^+$ divergence
is an artifact of the LF approach. It turns out that the kinetic mass
counter-term thus obtained is identical to the one determined by imposing
parity invariance for physical observables. In a non-perturbative
calculation, one obtains tachyonic behavior if the kinetic mass counter-term 
is smaller than its correct value. Above its correct value no tachyonic
behavior is observed, but the spectrum may or may not diverge if the
kinetic mass is too large. 
This ``critical'' behavior at the correct value can
be used as a signature for non-perturbative determinations of the kinetic
mass counter-term.

Unfortunately, there are several obstacles before one can apply
``finiteness conditions'' in practical calculations --- particularly in DLCQ. 
One reason is that the divergences that one needs to look for are only
logarithmic, which makes them hard to detect numerically. Furthermore, the 
situation in DLCQ is not quite as simple as it is in the continuum. DLCQ 
breaks manifest boost invariance, and the results in this paper show that
a simple Ansatz, where the kinetic mass counter-term is {\it not} a 
function of the momentum, is inconsistent in DLCQ already for perturbative
calculations within a super-renormalizable model. However, it is conceivable 
that a DLCQ calculation with additional cutoffs (such that momentum 
independence
of the results is achieved) can be based on Hamiltonians with momentum
independent mass counter-terms. These counter-terms can then, 
at least in principle,
be determined using the finiteness condition that was derived in this paper.

The results in this paper were based on perturbatively
analyzing Yukawa type interactions
in 1+1 dimensions,  and on non-perturbative
results involving fermions coupled to the $\perp$ component of a vector
field in 3+1 dimensions. 
It would be interesting to know what these results
imply for QCD in 3+1 dimensions. First of all, the limitation to 1+1 dimensions
can be easily overcome by introducing a lattice in the transverse space
coordinates. That way one obtains a 3+1 dimensional theory which is formally
equivalent to coupled 1+1 dimensional theories and the results of this
paper immediately translate. The real limitation of the results in this
paper is that while QCD contains interactions (couplings to the transverse
components of the gauge field) which resemble Yukawa
interactions, QCD also contains also couplings to the longitudinal
components of the gauge fields and those are much more singular for
$p^+ \rightarrow 0$ than the Yukawa-type couplings. It is not clear
whether renormalizing the kinetic mass will be sufficient to compensate
divergences arising from the couplings to the longitudinal components of
the gauge field as well. However, while it is not clear whether independent
renormalization of the kinetic mass will be  sufficient in QCD (most likely 
it is not), the mere fact that QCD contains interactions which resemble Yukawa 
interactions means that kinetic mass renormalization will be necessary.
Another result of this paper, namely that using only a DLCQ regulator is
inconsistent with a momentum independent mass term translates to QCD as
well. This comment also applies to dimensionally reduced models for QCD 
\cite{anton}.
\acknowledgements
I would like to thank Simon Dalley for useful discussions.
This work was supported by the D.O.E. (grant no. DE-FG03-96ER40965)
and in part by TJNAF.

\end{document}